\documentclass[aps,superscriptaddress,eqsecnum,nofootinbib,preprintnumbers]{revtex4}
\begin{document}
\def\GB{{\hat{\cal{G}}}}
\newcommand{\RR}{(*R*)}
\def\third{\textstyle{1\over3}}
\def\quarter{\textstyle{1\over4}}
\def\del{\partial}
\newcommand{\beq} {\begin{equation}}
\newcommand{\eeq} {\end{equation}}
\newcommand{\bea} {\begin{eqnarray}}
\newcommand{\eea} {\end{eqnarray}}
\newcommand{\bsea} {\begin{subeqnarray}}
\newcommand{\esea} {\end{subeqnarray}}
\newcommand{\nn}  {\nonumber}
\newcommand{\ga}{\gamma}
\newcommand{\ba}{\beta}
\newcommand{\lp}{\left(}
\newcommand{\rp}{\right)}
\newcommand{\dd}{\mathrm{d}}
\newcommand{\p}{\partial}
\newcommand{\da}{\delta}
\newcommand{\al}{\alpha}
\newcommand{\la}{\lambda}
\newcommand{\eps}{\epsilon}
\newcommand{\mc}[1]{\mathcal{#1}}
\newcommand{\half}{\frac12}
\newcommand{\ka}{\kappa}
\newcommand{\sig}{\sigma}
\newcommand{\U}{\hat U}
\renewcommand{\l}[1]{\left#1}
\renewcommand{\r}[1]{\right#1}
\renewcommand{\d}[1]{\displaystyle#1}
\newcommand{\ud}{\mathrm{d}}
\newcommand{\si}{ \mathrm{si}}
\newcommand{\co}{ \mathrm{co}}
\newcommand{\ep}{\epsilon}
\newcommand{\varep}{\varepsilon}
\newcommand{\Ga}{\Gamma}
\newcommand{\Om}{\Omega}
\newcommand{\La}{\Lambda}
\newcommand{\Th}{\Theta}
\newcommand{\Si}{\Sigma}
\newcommand{\G}{\mathcal G}
\newcommand{\F}{\mathcal F}
\newcommand{\A}{\mathcal A}
\newcommand{\ha}{\hat \alpha}
\newcommand{\ta}{\tilde \alpha}
\newcommand{\tar}{\tilde \alpha_r}
\newcommand{\mf}{\mathfrak m}
\newcommand{\qf}{\mathfrak q}
\newcommand{\V}{\mathcal V}
\newcommand{\bk}{\bar k}
\newcommand{\tk}{\tilde k}
\newcommand{\xxx}[1]{\noindent{\textcolor{red}{ \textbf{[#1]}}}}
\newcommand{\myref}[1]{{\tiny [\texttt{#1}]}}

\title{Self tuning scalar tensor black holes}

\author{Christos Charmousis}
\email{christos.charmousis@th.u-psud.fr}
\affiliation{Laboratoire de Physique Th\'eorique (LPT), Univ. Paris-Sud, CNRS UMR 8627, F-91405 Orsay, France}
\author{Damianos Iosifidis}
\email{diosifid@auth.gr}
\affiliation{Aristotle university of Thessaloniki, 541 24, Thessaloniki, Greece}

\date{\today}
\begin{abstract}

Studying a certain sub class of higher order Horndeski (scalar-tensor) theories we discuss a method discovered  recently permitting analytic black hole solutions with a non trivial and regular scalar field. One of the solutions found has de Sitter asymptotics and self tunes the bulk cosmological constant. Using the aforementioned method we find and analyse new black hole solutions which can also have de Sitter asymptotics. By looking at small deviations of the integration constant responsible for self tuning we discuss the robustness of the self tuning mechanism. We find that neighboring solutions to the one previously found present also self tuning properties with unaltered effective cosmological constant. 

\end{abstract}
\preprint{LPT-Orsay-15-05}
\maketitle

\section{Introduction-Motivation}

General Relativity (GR) is a theoretically and experimentally consistent theory. It is also mathematically unique. Indeed theoretical consistency and mathematical uniqueness emanate as a corollary to Lovelock's theorem{\footnote{The four dimensional version of this theorem was already known early on \cite{Cartan} and mathematically stems from the mathematical works of Chern \cite{Chern}}} \cite{chalovelock} in 4 dimensions: consider a smooth Lagrangian depending up to second order derivatives of the metric tensor $\mathcal{L}=\mathcal{L}(\mathcal{M},g,\nabla g, \nabla \nabla g)$ and endowed with a Levi Civita connection. Then in $D=4$ dimensions GR with cosmological constant  is the unique metric theory emerging from the action,
 \beq
 \label{unique}
	S_{\left( 4 \right)}  = \int_\mathcal{M} {d^4 x\sqrt { - g^{\left( 4 \right)} } \left[ {R - 2\Lambda} \right]} \nn
\eeq
which upon metric variation yields,
		\begin{itemize}
                        \item Equations of motion which are 2$^{\textrm{nd}}$-order PDE's,
			\item given by a symmetric two-tensor, $G_{\mu\nu}+\Lambda g_{\mu\nu}$
			\item and admitting Bianchi identities.
		\end{itemize}
Under these hypotheses GR with a cosmological constant is the unique massless, spin 2,  4 dimensional theory of gravity! In other words the Einstein-Hilbert action is the unique metric action which evades  Ostrogradski no-go theorem of 1850 \cite{chaostro}. This no-go theorem states that any non-degenerate, higher than second order derivative theories inevitably lead to ghost degrees of freedom! In fact the only safe metric term we can add to the above action is the Gauss-Bonnet scalar, $R^2-4 Ricci^2+Riemann^2$, whose action is a 4 dimensional topological invariant the generalized Euler characteristic. Although this geometric term is not trivial its metric variation is not modifying thus GR equations of motion.

GR is a physical theory which has remarkable agreement with weak and strong gravity experiments from scales smaller than the millimeter and up to 30 times Earth to Sun distance...
At astrophysical scales however, GR equations of motion, require dark matter interacting only gravitationally. The presence of dark matter although well founded theoretically still evades astroparticle and accelerator experimental detection. Furthermore, at cosmological scales the Universe is now observed to be in an accelerating phase requiring some source of yet unknown dark energy in order for Einstein's equations to agree with a multitude of differing cosmological and astrophysical observations \cite{chasuper}. In resum\'e, if we assume only ordinary (detected) sources of matter there is a flagrant disagreement between local, astrophysical and cosmological data. In fact roughly about 4\% is matter of known source in the Universe. Even if we accept theoretically the presence of dark matter, dark energy of about 70\% still has to be accounted for. For dark energy, which we will focus from here onwards, there is an almost trivial way out. Assume a cosmological constant  of energy density $\rho_\Lambda=\frac{\Lambda_{obs}}{8\pi G}=(10^{-3}eV)^4$, ie. modify Einstein's equation by,
\beq
\nn
G_{\mu\nu}+\Lambda_{obs} g_{\mu\nu}=8\pi G T_{\mu\nu}
\eeq
Now this is a rather economic and consistent solution since a cosmological constant is the only other metric term allowed  given the above uniqueness theorem (\ref{unique}). However this cosmological scale is tiny, the mass scale corresponds to neutrino mass scales. This is because a cosmological constant introduces a scale in the metric, $r_0\sim \sqrt{\Lambda}$ which is as tiny, as the inverse size of the  Universe today, $r_0=H_0^{-1}$. Note that the quotient made out of Solar system scales divided by Cosmological Scales leads to an enormous number $\frac{10 A.U.}{H_0^{-1}}=10^{-15}$ at which we are probing the physical theory in question, GR, compared to the local experimentally robust scales. In fact suppose we were to use local experiments in order to bound the cosmological constant value (without taking into account cosmological data). That would involve constraining the cosmological constant  from the (very) well known trajectories of planets. In essence, replacing Schwarzschild geometry by de Sitter Schwarzschild and taking the PPN limit. Local experiments would allow  a cosmological constant roughly $10^{10}$ bigger than its current cosmological bound! Physical theories describe the physical world from good to remarkable precision but up to (or at) certain scales. Are we therefore beyond the scale of application of GR? Although deep in the UV, i.e. close to the Planck scale, GR is ill defined, could it be that also at vast and extremely dilute cosmological scales, of almost zero curvature, some other theory completes the theory of General Relativity? Furthermore, the cosmological constant is accompanied by the cosmological constant problem: its huge theoretical value, as expected from QFT considerations is at complete contrast with a finely tuned and tiny cosmological constant  \cite{chaweinberg} (for a complete discussion and interesting developments see also \cite{chakaloper}).

Summarizing: what if the need for exotic matter or a tiny cosmological constant is the sign for novel gravitational physics at very low energy scales or large distances. It is clearly a valid avenue of theoretical research in order to understand the robustness of GR at $10^{15}$ greater than local distances.
In fact historically there is a very similar paradigm encountered at the advent of GR. More than 100 years ago, there were also observational short comings of Newton's theory of gravity. Indeed, Mercury the closest orbiting planet around the sun was deviating from Keplerian motion by a tiny yet definite advance in its perihelion. At the time, differing possibilities were pursued including most interestingly, the existence of a novel planet, Vulcan, which was postulated to orbit even closer to the sun, slightly deviating the perihelion of Mercury much like a dark source to Newtonnian equations.
In fact a simple next order correction with one additional parameter (much like a cosmological constant) was enough to save Newton's laws (at the experimental precision of the time). Furthermore, although the presence of Vulcan may seem now days an ad hoc hypothesis it actually was not. The outermost planet of the solar system, Neptune, was in fact first theoretically predicted prior to being observed by the (same) astronomer Le Verrier. Indeed the theoretical prediction was triggered due to the reported inconsistencies in the Newtonian planetary trajectory of Uranus-which were explained by the presence of Neptune (see the exciting account in \cite{Ferreira})!
For the advance of the perihelion of Mercury, however, it was not the "simpler" solution that was the correct one, indeed GR, a far more involved than Newtonian theory had to be employed in order to explain strong gravity at local scales as was accepted by the scientific community only later by Eddington s famous eclipse experiment. 
Just as the eventual success of GR was not explaining the advance of Mercury's perihelion, modification of gravity cannot only be "an explanation" of  the  effective cosmological constant. Successes of GR underlies in all the theoretical and subsequently observational advances such as, the Big Bang Universe, inflation theory, black holes, binary pulsars, quasars etc. 

So which avenues should one choose in an effort to modify such a well established theory, both theoretically and experimentally? For a start, we can deduce a set of basic facts/guidelines that can serve as a rough guide to possible research directions:
\begin{itemize}
\item Since GR is the unique 4 dimensional metric theory we will need to introduce new and genuine gravitational degrees of freedom!
\item Novel and non degenerate gravitational degrees of freedom must not lead to higher derivative equations of motion. For then additional degrees of freedom are ghosts and vacuum is unstable due to Ostrogradski theorem 1850 as explained in \cite{woodard}. 
\item Matter must not directly couple to novel gravity degrees of freedom. Matter sees only the metric and evolves in metric geodesics. As such EEP is preserved and space-time can be put locally in an inertial frame. 
\item Novel degrees of freedom need to be screened from local gravity experiments. We need a well defined GR local limit such as the Chameleon, see for example\cite{Khoury}, or the Vainshtein mechanism \cite{Babichev}. The equivalent of this in GR would be the weak gravity Newtonian limit.
\item Exact gravitational solutions are essential in modified gravity theories in order to understand strong gravity regimes and novel characteristics. We will need to deal with the no hair paradigm and the absence of GR theorems such as Birkhoff's theorem in modified gravity theories. 
\item A modified gravity theory should tell us something about the short-comings of GR such as the cosmological constant problem and in particular how to screen an a priori enormous cosmological constant.
\end{itemize}
The above list of points are not exhaustive and one must treat them not dogmatically but rather as rough guidelines. For example f(R) theory has higher order derivative equations of motion but does not necessarily acquire ghost degrees of freedom. The theory can be easily transformed in a scalar-tensor theory. The spin 2 degrees of freedom are in fact not present for this (degenerate) case but only an extra scalar degree of freedom. Having this  in mind we can sketch, again with precaution, differing directions to go beyond GR (for a full account see for example \cite{chapadilla}):
\begin{itemize}
\item We can assume metric theories but in the presence of extra dimensions : Extension of GR to Lovelock theory with modified yet second order field equations \cite{chareviews}. Here belong braneworld models  such as the DGP model \cite{DGP}, RS models \cite{RS} or the generic case of Kaluza-Klein compactification.
\item One can assume that the graviton is not massless but massive. In fact in this case generically graviton potentials suffer from a Boulware Deser ghost mode \cite{Boulware:1973my}. The only possible theories avoiding this specific mode are non trivial and are parametrized by 3 independent couplings (see for example \cite{chamassive}). Hence massive gravity and bigravity theories are greatly constrained already theoretically and the most general interaction (potential) term is known.
\item One can assume a 4-dimensional scalar modification of GR. This is the case of Scalar-tensor theories \cite{chaBD}, \cite{chafarese}, $f(R)$ \cite{chasotiriou}, Galileon/Hornedski theories \cite{chahorndeski}, \cite{chanicolis0},\cite{chadeffayet} (for a review and relation to Lovelock theory see \cite{Charmousis:2014mia}).
\item We can consider breaking symmetries present in GR such as Lorentz invariance symmetry. This has interesting implications in the UV as it is a way of obtaining higher order theories which are power counting renormalizable and have a priori better ultra-violet properties: this is the case of Horava gravity \cite{Horava}, \cite{Blas} and in its infra red limit, Einstein Aether theories \cite{Jacobson},  which are healthy and unique vector tensor theories with a consistent GR limit.
\item Theories modifying geometric properties of spacetime such as with the inclusion of torsion or a different choice of geometric connection (see for example \cite{chaolmo}).
\end{itemize}
In this paper we will consider the simpler of modifications with a single extra degree of freedom, scalar-tensor theory. Scalar tensor theories are in a certain sense a prototype of gravity modification and have some nice properties. In particular,
\begin{itemize}
\item They admit a uniqueness theorem due to Horndeski, 1973, which much like Lovelock's theorem determines the most general form of their action \cite{chahorndeski}, \cite{chadeffayet}. 
\item They contain or are limits of other modified gravity theories. $f(R)$ theory for example is a scalar tensor theory in disguise whereas massive gravity or bigravity are scalar tensor theories in their decoupling limit. This is also true of certain higher dimensional braneworlds such as DGP and even for Kaluza-Klein compactifications of Loveock theory \cite{VanAcoleyen:2011mj}, \cite{chablaise}, \cite{Charmousis:2014mia}
\item They can have insightful screening mechanisms as in the Chameleon \cite{Khoury} and Vainshtein \cite{Babichev} mechanisms.
\item They include specific Galileon terms, dubbed Fab 4, that can screen classically a big cosmological constant \cite{chafab4}.
\item They specifically break GR black hole uniqueness theorems such as Birkhoff's theorem and suffer from no-hair black hole theorems. 
\end{itemize}
It is this latter characteristic we will study and try to overcome for we will seek black hole solutions of such theories where the scalar field will not be trivial but well behaved. Before doing so let us point out the most general theory at hand. 
Horndeski showed several years ago with techniques similar to those of Lovelock that the most general scalar-tensor action with second order field equations is given by
\begin{eqnarray}
\label{hdaction}
S_H&=&\int d^4 x \sqrt{-g}\left(L_2+L_3+L_4+L_5\right)\, \nn
\end{eqnarray}
\bea
L_2 &=& K(\phi,X)   \nn  ,
\\
L_3 &=& -G_3(\phi,X) \nabla^2 \phi,\nn
\\
L_4 &=& G_4(\phi,X) R + G_{4X} \left[ (\nabla^2 \phi)^2 
-(\nabla_\mu\nabla_\nu\phi)^2 \right],\nn
\\
L_5 &=& G_5(\phi,X) G_{\mu\nu}\nabla^\mu \nabla^\nu \phi - 
 \frac{G_{5X}}{6} \left[ (\nabla^2 \phi)^3 - 3\nabla^2 
\phi(\nabla_\mu\nabla_\nu\phi)^2 + 2(\nabla_\mu\nabla_\nu\phi)^3 \right] \, \nn
\eea
where the $G_i$ are unspecified functions of $\phi$ and $X\equiv-\frac{1}{2} \nabla^\mu \phi \nabla_\mu \phi$ and  $G_{iX}\equiv \partial G_i/\partial X$. 
In fact this is the action of covariant Galileons found independently in \cite{chaDeffayet:2009wt} in a different and elegant manner.  Due to the higher order kinetic terms in $X$ the theory generically screens the scalar mode locally following the Vainshtein mechanism \cite{Babichev}.  

In the next section we will discuss scalar tensor black holes starting with a brief discussion of the no hair paradigm, and then certain solutions some of which are new and appear in the last subsection. We will then briefly conclude.

\section{Scalar tensor black holes}

Modified gravity theories have to deal with the absence of black hole uniqueness theorems (present in GR) as well as no hair theorems. Indeed, "Black holes have no hair", is the celebrated phrase conjectured long ago by Wheeler. The no hair paradigm takes concrete form as a set of mathematical theorems governing asymptotically flat (or de Sitter) and stationary black holes under certain hypotheses. The conjecture itself states, that apart from charges measured at infinity by a far away observer, no additional degrees of freedom can describe the black hole geometry (for a review see \cite{bekenstein}). The  physical picture is roughly the following: as black holes are formed  they expel or eat up surrounding matter ending up, in their stationary phase, as rather "simple" bald objects characterized by a finite number of measurable charges at infinity: mass, angular momentum, electric and magnetic charge. A black hole, is a rather blunt gravitational object having specific charges and not allowing additional parameters Ð primary hair Ð which are not associated to a conserved charge, or, secondary hair of no additional parameters but non-trivial fields interacting with the black hole spacetime. An ideal observer measuring these charges out at infinity "knows" specifically which black hole she or he is dealing with. They do not have to venture close up in order to gather details about the black hole structure, its horizon properties etc. This is true for asymptotically flat and stationary black holes under certain hypotheses{\footnote{Recently an interesting  counter-example  to black hole no hair theorems was found for stationary black holes with a massive complex scalar field. There crucially the scalar field was assumed not to have the same symmetries as the spacetime metric \cite{Herdeiro}.}}. As such, any scalar field, part of a scalar tensor theory is expected to be frozen to a constant in order to evade a naked singularity. In other words scalar-tensor theories in their prototype Brans Dicke form \cite{chaBD} basically have the same GR black holes with the scalar field fixed to a constant value. As such a spherically symmetric BD black hole is again a Schwarzschild black hole with a constant scalar field. Otherwise the scalar tensor spherically symmetric solutions are singular geometries describing rather stars which in BD theory are different solutions to those describing black holes. We emphasize here the difference with GR where the Schwarzschild solution is the exterior metric for a star and a black hole due to Birkhoff's uniqueness theorem. This is not true in scalar tensor theories and it is a source of problems of scalar tensor theories with local experiments. There is a non-trivial example of a black hole geometry that almost escapes this rule and that is the BBMB black hole which we will briefly turn to now.

\subsection{An example: The BBMB black hole}

Consider the following scalar tensor action,
\beq
\label{BBMB}
S[g_{\mu\nu},\phi,\psi] = \int_\mc M \sqrt{-g} \lp \frac{R}{16\pi G} - \frac12 \p_\al \phi \p^\al \phi - \frac1{12}R\phi^2 \rp \dd^4 x
\eeq
The particular feature of this action is that the scalar $\phi$  couples conformally to the metric. 
Indeed there is an invariance of the $\phi$ equation of motion (EoM) under the conformal transformation
\bea
 g_{\al\beta} &\mapsto \tilde{g}_{\al\beta} = \Om^2 g_{\al\beta} \\
\phi &\mapsto \tilde{\phi} = \Om^{-1} \phi 
\eea 
The whole action is not conformally invariant.  In fact the Einstein-Hilbert term explicitly breaks conformal invariance. However, for this action for which there is a remnant of conformal symmetry  there exists a vacuum black hole geometry with non-trivial scalar field and secondary black hole hair.
This is the BBMB solution \cite{BBMB} a static and spherically symmetric solution,
$$ \dd s^2 = - \lp 1-\frac{m}{r} \rp^2 \dd t^2+ \frac{\dd r^2}{\lp 1-\frac{m}{r} \rp^2} + r^2 \lp \dd\theta^2 + \sin^2\theta \dd\varphi^2 \rp $$
with a non trivial scalar field
$$\phi = \sqrt{\frac{3}{4\pi G}} \frac{m}{r-m} $$
This particular geometry is also encountered in GR. It is that of an extremal Reisner-Nordtstrom solution with charge equal to mass. Our ideal observer sitting far out of this scalar tensor black hole would observe a static black hole of mass $m$ other than Schwarzschild but with no electric charge. Indeed note that the scalar decays at infinity and does not carry a specific scalar charge. Hence we are not dealing  with primary hair (by definition) rather secondary hair as the scalar couples non-trivially (conformally) to the metric modifying its geometry. Indeed the only parameter of the solution is the black hole mass $m$.  Almost immediately however we encounter the shortcoming of the solution: the scalar field explodes at the horizon location at $r=m$. This can be remedied in the presence of a cosmological constant \cite{Martinez} and a $\phi^4$ conformal potential. The black hole can be even in an accelerated phase or have a Taub NUT charge \cite{cmetric}. Here, for the asymptotically flat case, the singular scalar at the horizon is a problem and it is not clear whether the solution is a black hole i.e. an endpoint of matter collapse. Going to the Einstein frame we can actually see that the solution hides additional singularities there at $r=2m$  where the conformal transformation in between the frames is singular. In fact at $r=2m$ the $\phi$ dependent factor multiplying $R$ in the action is exactly zero and for $r<2m$ changes sign hinting on gravity acquiring a ghost kinetic term! Could self gravitating matter go beyond the $r=2m$ surface? Or again, is the BBMB solution an endpoint of gravitational collapse? It would seem unlikely and this is an open problem which could be resolved numerically determining the exact nature of this solution. The BBMB metric is however a unique and simple 4 dimensional example with a non trivial scalar field. We will proceed now to remedy at least some of these problems with higher order Galileon fields.

\section{Constructing Galileon black holes}

We will now consider the action{\footnote{The details summarized in this section can be found in the original paper \cite{Babichev:2013cya}. See also \cite{max} for the first black hole solutions discussed for this theory}},
\begin{equation}\label{action}
S = \int d^4x \sqrt{-g}\left[\zeta R  - 2 \Lambda -\eta \left(\partial\phi\right)^2 +\beta G^{\mu\nu}\partial_\mu\phi \partial_\nu\phi  \right],
\end{equation}
which is part of the Horndeski action given above (\ref{hdaction}) comprising of the Einstein-Hibert term with a cosmological constant as well as the canonical scalar kinetic term and a higher order scalar tensor interaction term. The latter term is a higher order Galileon, central in our forthcoming discussion, yielding second order field equations. This nice property boils down to the divergence freedom of the Einstein tensor. Finally $\zeta$, $\eta$ and $\beta$ are coupling constants in order to keep track of the terms in the action. Indeed the metric field equations are,
\bea
\zeta G_{\mu\nu} &-\eta \left(\partial_\mu\phi \partial_\nu\phi -\frac12 g_{\mu\nu}(\partial\phi)^2 \right)  +g_{\mu\nu}\Lambda \\
	&+\frac{\beta}2 \left( (\partial\phi)^2G_{\mu\nu} + 2 P_{\mu\alpha\nu\beta} \nabla^\alpha\phi \nabla^\beta\phi \right. \\
	 & \left.   +  g_{\mu\alpha}\delta^{\alpha\rho\sigma}_{\nu\gamma\delta}\nabla^\gamma\nabla_\rho\phi \nabla^\delta\nabla_\sigma\phi \right)
	=0,
\eea
where $P_{\mu\alpha\nu\beta}$ is the double dual curvature tensor given by,
\beq
P_{\mu \rho \nu \sigma}=R_{\mu\rho\nu\sigma}+R_{\nu\rho}g_{\mu\sigma}+R_{\mu\sigma}g_{\nu\rho}-R_{\rho\sigma}g_{\mu\nu}-R_{\mu\nu}g_{\rho\sigma}+{1\over 2}R(g_{\mu\nu}g_{\rho\sigma}-g_{\mu\sigma}g_{\nu\rho}).
\eeq
This tensor is divergence free has the same symmetry properties as the Riemann tensor and tracing two of its indices gives,
$$
g^{\mu\nu}P_{\mu\rho\nu\sigma}=-G_{\rho\sigma}
$$
the Einstein tensor.

An important property of the above action is that the scalar field has translational invariance: $\phi\to \phi +$const. This is important for then the scalar field equation can be written in terms of a conserved vector current,
\begin{equation}\label{eomJ}
 \nabla_\mu J^\mu =0,\;\; J^\mu = \left( \eta g^{\mu\nu} -\beta G^{\mu\nu} \right) \partial_\nu\phi.\nn
\end{equation}
Note in the last expression of the current above, the explicit appearance of the metric field equations of the lower order Einstein-Hilbert and cosmological constant terms. Schematically we have that, $R \rightarrow G^{\mu\nu}\partial_\mu\phi \partial_\nu\phi$ whereas $\Lambda\rightarrow g^{\mu\nu}\partial_\mu\phi \partial_\nu\phi$ associating thus in this intriguing way the terms present in the action. Summarizing therefore, apart from the latter property, the above action crucially includes higher order Galileon terms and has translational invariance for the scalar field.

We will consider a static and spherically symmetric spacetime, 
\beq
ds^2 = -h(r)dt^2 + \frac{dr^2}{f(r)} + r^2 d\Omega^2
\eeq
parametrized by the two radial functions $f$ and $g$. It is then evident that if $\phi=\phi(r)$ then the scalar equation is integrable and reads,
\beq
(\eta g^{rr} -\beta G^{rr}) \sqrt{g} \phi'=c
\eeq
where $c$ the integration constant stands for primary hair for a scalar-tensor black hole. A non zero value of $c$ comes along with regularity problems
since the current norm is singular $J^2=J^\mu J^\nu g_{\mu\nu}=(J^r)^2 g_{rr}$ unless $J^r=0$ at the horizon and hence $c=0$.
It is then easy to note that generically $\phi=constant$ everywhere as noted in \cite{chanicolis} and we have the appearance of a no-hair argument i.e. the presence of a trivial scalar and a GR black hole solution. However, here there is a slight twist in this argument. Indeed, due to the higher order nature of the terms in the action we can instead of $\phi=constant$ require that,
\beq
\label{damos}
\beta G^{rr}-\eta g^{rr}=0.
\eeq
and thus search for a different branch of solutions. Here note that the presence of the Einstein tensor (\ref{damos}) in the higher order term, is just the Hamiltonian constraint for a GR plus cosmological constant theory, and this is the reason for our Gallileon choice. A second point is that we can also include a linear time dependence in the scalar field  $\phi(t,r) = q\, t + \psi(r)$ without affecting the scalar field equation, 
$-\partial_r[(\beta G^{rr}-\eta g^{rr})\partial_r \psi ]-\partial_t[(\beta G^{tt}-\eta g^{tt})\partial_t (q t) ]=0$ which is still verified given (\ref{damos}). In fact, we can show that choosing $\phi(t,r) = q\, t + \psi(r)$ solves also the $(tr)$-equation of motion and this at the end tells us that our Anzatz is mathematically consistent. This interesting property where the scalar field does not have the same symmetries as the background metric while having no flux is very similar to the idea used in \cite{Herdeiro} to construct stationary minimally coupled scalar hair. We must stress however that although the scalar is not necessarily constant we have no scalar charge i.e. $c=0$. Furthermore we have a finite current. We have satisfied the no hair argument not in permitting a non trivial scalar charge in the face of the constant $c$, but still having a scalar field which is not trivial. This is due to the higher order nature of the Galileon term providing us with an alternative geometric resolution $\beta G^{rr}-\eta g^{rr}=0$ to the scalar and $tr$-equation. Therefore the $tr$-equation as well as the scalar field equation boil down to the geometric constraint, 
$$f = \frac{ (\beta+\eta  r^2) h}{ \beta (rh)'},$$ fixing the spherically symmetric gauge and $\phi(t,r) = q\, t + \psi(r)$.

This solves therefore for the time dependence of $\phi$ and for the gauge function $f$.
In order to find solutions we now need to solve for $\psi(r)$ and $h(r)$ with the last two independent ODE's, the $(rr)$ and $(tt)$ equations. 
From the (rr)-component we get $\psi'$
\begin{equation}\label{psi}
	\psi' = \pm \frac{\sqrt{r}}{h(\beta+\eta r^2)}\left(q^2\beta (\beta+\eta r^2) h'-\frac{
	\lambda}{2}(h^2r^2)'\right)^{1/2}.\nn
\end{equation}
with $$\lambda\equiv \zeta\eta + \beta\Lambda$$. 
We see that for $\eta=\Lambda=0$ time dependence of the scalar is essential in order to have a non trivial asymptotically flat solution.
Finally the (tt)-component gives $h(r)$ via,
\begin{equation}\label{h}
	h(r) = -\frac{\mu}{r} +\frac{1}{r}\int \frac{k(r)}{\beta+\eta r^2}dr,\nn
\end{equation}
with
\begin{equation}\label{k}
	q^2\beta (\beta+\eta r^2)^2 - \left(2\zeta\beta+\left( 2\zeta\eta -\lambda \right) r^2\right) k + C_0 k^{3/2} =0,\nn
\end{equation}
Therefore any solution to the algebraic eq for $k=k(r)$ gives us a full solution to the system. For $\eta=\Lambda=0$ we can easily find asymptotically flat solutions with a scalar field since then $k$ is simply a constant which can be gauged away in a redefinition of time. The geometry is that of a Schwarzschild black hole \cite{Babichev:2013cya} and the scalar field due to the linear time dependence is regular at the black hole future horizon. The scalar field  linearly diverges at infinity but note that its derivatives appearing in the action are well behaved.  The solution is therefore a valid and regular solution unlike the BBMB case where the scalar blows up at the horizon location. 

One can then seek asymptotically de-Sitter solutions \cite{Babichev:2013cya} setting $f=h$ whereupon 
\beq
\label{desitter} 
k(r)=\frac{(\beta+\eta r^2)^2}{\beta}
\eeq
 with
 \beq
 \label{st}
 q^2 = (\zeta\eta + \beta\Lambda)/(\beta\eta), \qquad C_0 = (\zeta\eta-\beta\Lambda)\sqrt{\beta}/\eta.
\eeq
  The line element is that of de Sitter Schwarzschild $f  =h = 1- \frac{\mu}{r} + \frac{\eta}{3\beta}r^2$ with $\psi' = \pm\frac{q}{h}\sqrt{1-h}$ and $\phi(t,r) = q\, t + \psi(r)$. Going to Eddington-Filkestein coordinates we can check that the solution is regular at the horizon. Note again that the scalar itself diverges at the de Sitter horizon but not its derivatives. 

The solution above has an interesting property : it is a GR like solution but crucially it does not directly depend on the vacuum cosmological constant $\Lambda$ but rather on an effective cosmological constant $\Lambda_{eff} = - \eta/\beta$. This is true as long as the integration constant $q^2 \eta= \Lambda-\Lambda_{eff}>0$ which is just (\ref{st}). Also note that the solution belongs to a different branch than the GR solution of de Sitter Schwarzschild, one cannot switch off the higher order Gallileon term. Hence any arbitrary bulk cosmological constant of the action $\Lambda>\Lambda_{eff}$, which can be as big as the Planck scale, fixes $q$ our integration constant leaving us with an effective geometric acceleration $\Lambda_{eff}$ which can be tuned once and for all in the theory to be small. In other words the scalar tensor solution self tunes the vacuum cosmological constant as long as $\Lambda>\Lambda_{eff}$. But what is the behavior of other solutions to the algebraic equation (\ref{k})? In particular is there a fine tuning in the integration constant q? If we deviate from the conditions (\ref{st}) will we still have de Sitter Schwarzschild asymptotics? We will turn to this question now.

\section{General Solutions}

In the previous section we summarized the method and certain key solutions of spherical symmetry found in \cite{Babichev:2013cya}. We would now like to go further and study more generic solutions of the algebraic master equation (\ref{k})  governing the general solution of the system.  The first important thing to note is that the solution of the cubic $k=k(r)$, as given by (\ref{k}) does not depend on the mass integration constant $\mu$ of the solution (\ref{h}). The mass of the black hole solution does not interfere in the equation itself. The algebraic equation (\ref{k}) specifies the asymptotic behavior of the solution via the integration constant $C_0$. As such, solution (\ref{desitter}) corresponds to de Sitter space with a non trivial scalar field $\phi$ and can be promoted to a black hole for $\mu\neq 0$. We will refer to it as a de Sitter solution from now on (see in that aspect \cite{Gubitosi:2011sg}). What other solutions $k=k(r)$ of de Sitter asymptotics exist other than (\ref{desitter})? It is easy to verify that asymptotically (anti) de Sitter solutions will be obtained by fixing the constant $C_0$ to its found de Sitter value given above for the de Sitter Schwarzschild geometry (\ref{desitter}). The fake hair charge $q$ remains a free parameter. We will look into these solutions in order to understand their behaviour beyond the GR stealth limit (\ref{desitter}) and in particular to see locally around (\ref{desitter}) how self tuning is influenced in this class of solutions. 

But first of all there is also a second class of solutions we can obtain relatively easily. They are simply given by dividing the third order polynomial by the de Sitter root (\ref{desitter}). The resulting quotient polynomial will yield  a second degree equation for $k(r)$ which can have real roots and which can be easily solved.  Indeed we know that the parameter choice $q^{2}=\frac{\zeta \eta+\beta\Lambda}{\beta\eta}$ and $C_{0}=(2\zeta\eta-\lambda)\sqrt{\beta}/\eta$ yields (\ref{desitter}) as a solution of (\ref{k}). 

In what follows we will consider the cases $\beta<0$ and $\beta>0$ separately. For the second order equation it is $\beta>0$ that presents more interest in terms of regular solutions. For the former de Sitter solutions they will be in the same family as (\ref{desitter}) hence $\beta<0$. Now that we have understood the influence of each term of the action in the field equations we will set natural values for the couplings in the action $\zeta=1$, $\eta=1/2$ and consider $\Lambda>0$ while defining $R=\frac{r}{\sqrt{2|\beta|}}$ and $u=2|\beta|\Lambda$ without loss of generality. Hence the $u$ parameter is dimensionless and denotes the strength of the higher order coupling $\beta$ for given cosmological constant. 

For $\beta>0$ we have $q^{2}|\beta|=(1+u)$, $C_{0}=\sqrt{|\beta|}(1-u)$ and setting
\begin{equation}
x=\frac{\sqrt{\beta}}{k^{1/2}}
\end{equation}
the cubic (\ref{k}) takes the canonical form,
\beq
\label{x1}
x^{3}- \frac{\Big[ 2+(1-u)R^{2} \Big]}{(1+u)(1+R^{2})^{2}} x +\frac{(1-u)}{(1+u)(1+R^{2})^{2}}=0
\eeq
Solution (\ref{desitter}) reads in the novel rescaled variables,
\begin{equation}
x=\frac{1}{1+R^{2}}
\end{equation}
So, writing again the cubic as
\begin{equation}
\Big( x-\frac{1}{1+R^{2}} \Big) (x^{2}+bx +c)=0
\end{equation}
we conclude that
\begin{equation}
b=\frac{1}{1+R^{2}}>0
\end{equation}
while
\begin{equation}
c=-v \frac{1}{1+R^{2}}=-v b
\end{equation}
where we have set for simplicity
\beq
\label{v}
v=\frac{1-u}{1+u}.
\eeq
It is then easy to show that the remaining two solutions of (\ref{x1}) are given by
\begin{equation}
x_{1,2}(R)=\frac{1}{2(1+ R^{2})}\left[ -1\pm \sqrt{1+4 v (1+ R^{2})} \right]
\end{equation}
For $v>0$ both of the above roots are real which is true for $-1<2\Lambda \beta<1$ 
and the solution avoids branch singularities (where the square root is zero). Computing the integral in (\ref{h}) we get, 
\begin{equation}
R \; h(r)=\sqrt{2\beta}\int \frac{k(r)}{\beta+ \eta r^{2}} dr=4\int \frac{(1+R^{2})}{\Big[- 1\pm \sqrt{1+4 v(1+R^{2})}\Big]^{2}} dR
\end{equation}
whereas 
\beq
f(R)=h(R) \Big[ -1\pm \sqrt{1+4 v (1+R^{2})}\Big]^{2}
\eeq
Therefore $f$ has at least the same zeros as the function $h$ which will be event horizons. At the end the solution for $h$ reads after direct integration,
\beq
\label{h1}
h(R)=\frac{1}{v} -\frac{\mu}{R}+\frac{1}{2 v^2 R}\left(\arctan(R)\pm \arctan\left(\frac{R}{\sqrt{1+4 v (R^2+1)}}\right)\right)\pm\frac{1}{v^{3/2} R}\sinh^{-1} \left(\sqrt{\frac{4vR^2}{1+4 v}}\right)
\eeq
Therefore, for large $R$, $h=h(R)$ is bounded and asymptotes $1/v$. For $\Lambda=0$ we have $v=1$ and this is the only case where the area of the sphere is not reduced. Otherwise we have a solid angle deficit which at the absence of an event horizon can lead to a naked singularity. 
For $\mu>0$ there is always an event horizon covering the singularity at $R=0$ and the above solution is a black hole with a well behaved scalar field. We have found therefore a regular scalar tensor black hole solution quite similar to the black hole embedded in an Einstein static universe \cite{Babichev:2013cya}. 

A second case of interest to consider is to take $C_{0}=(2\zeta\eta-\lambda)\sqrt{\beta}/\eta$ while allowing for arbitrary charge $q$. Our aim here is to seek solutions which asymptote de Sitter space and generalize the de-Sitter Schwarzschild solution found in \cite{Babichev:2013cya} given by (\ref{desitter}). So in accordance to this 
aim we take $\beta<0$, that is $\beta=-|\beta|$. Now, the constant $C_{0}$ takes the value
\begin{equation}
C_{0}=i\sqrt{|\beta|}(1+u)
\end{equation}
so that
\begin{equation}
k \sim \frac{\eta^{2}}{\beta} r^{4}=-\frac{\eta^{2}}{|\beta|} r^{4}
\end{equation}
as $r\rightarrow \infty $. Given the imaginary value of $C_0$, setting  $ix=\frac{\sqrt{|\beta|}}{k^{1/2}}$ the cubic becomes
\beq
x^{3} +\frac{\Big[ 2-(1+u)R^{2} \Big]}{q^{2}|\beta| (1-R^{2})^{2}}x +\frac{(1+u)}{q^{2}|\beta| (1-R^{2})^{2}}=0 \label{qw}
\eeq
Expressions are slightly more involved here and we make some redefinitions to render expressions more readable. Setting,
\beq
y=1-R^2,\qquad p^2=\frac{q^2 |\beta|}{u+1}
\eeq
the cubic reads,
\beq
\label{babi}
x^3+\frac{y+v}{p^2 y^2}x+\frac{1}{p^2 y^2}=0
\eeq
where $v$ is given in (\ref{v}) and we note that,
\beq
R \; h(R)=\int\frac{dR}{y(R) \, x_0^2(y)}, \qquad f(R)=h(R) y^2 x_0^2(y)
\eeq
where $x_0(y)$ is solution to the cubic (\ref{babi}).
Customary to third order equations we define
\begin{equation}
Q=\frac{1}{p^2 y^2}, \qquad P=(y+v)Q
\end{equation}

The number of real solutions of $(\ref{qw})$ depends upon the sign of the discriminant
\begin{equation}
\Delta= Q^{2}+4 \left(\frac{P}{3}\right)^{3}
\end{equation}
More specifically , for $\Delta>0$ we have only one real root and a complex pair. Since we are looking for a static patch of De Sitter we take $0<y<1$. Indeed the de Sitter Schwarzschild solution is obtained for $p^{2}=-v$ and $x_{dS}=\frac{1}{y}$. It is then clear that $\Delta>0$ and the unique real solution is given by
\begin{equation}
x_{0}=\frac{1}{2^{1/3}p y }\left[  -p y+\sqrt{p^2 y^2+\frac{4}{3^3} (y+v)^3} \right]^{1/3}   -\frac{2^{1/3} (y+v)}{3p y \left[  -p y+\sqrt{p^2 y^2+\frac{4}{3^3} (y+v)^3} \right]^{1/3}  }
\end{equation}

The explicit solution can be obtained by a simple but tedious procedure. Here we are mostly interested in determining the asymptotic behavior of the solution close to the de Sitter like horizon in $y=0$. Hence we consider the leading order behavior of the real solution(s) in small $y$ for differing values of $v$ respectively negative and positive. The stealth de-Sitter solution (\ref{desitter}) belongs to the family with $v<0$. In this case we obtain, $x_0\sim \frac{\sqrt{-v}}{p y}$ and this is again a de-Sitter Schwarzschild solution to leading order in $y$. Therefore, the de Sitter Schwarzschild solution is not a particularly special solution it is continuously related to a full branch of de Sitter like black hole solutions with similar characteristics. For $v>0$ we get however $x_0\sim \frac{1}{-v}$ which yields $R h=v^2 \tanh^{-1} R -\mu$ which has a singular behavior asymptotically. Therefore we see that for $v<0$ the solutions are indeed similar to (\ref{desitter}).

In order to back up this conclusion without going to numerics we can also consider the following approximation. Suppose that we have a small deviation from the charge configuration valid around the de-Sitter solution (\ref{desitter}), namely
\begin{equation}
q^{2}|\beta| = (u-1)+\epsilon
\end{equation}
with $\epsilon<<|u-1|$.
We consider the linearized solution in $\epsilon$ to ($\ref{qw}$), 
\begin{equation}
x=x_{0}+\epsilon x_{1}
\end{equation}
where $x_0=\frac{1}{R^2-1}$ and $x_{1}$ is the deviation from the de-Sitter solution (\ref{desitter}) when we shift slightly the fake hair parameter. Now, upon substituting the above into ($\ref{qw}$) to linear order in $\epsilon$ we get,
\begin{equation}
x(R)=\frac{1}{(1-R^{2})}\left[ 1 + \frac{\epsilon}{(3 u-1)-(1+u)R^{2}} \right].
\end{equation}

Now, if $u \in (-1,1/3)$ defining,
\begin{equation}
\frac{1+u}{1-3u}=\gamma^{2}>0
\end{equation}
we get
\begin{equation}
h(R)=\left( 1-\frac{2}{1+u}\epsilon \right)-\frac{\mu}{R}-\frac{R^2}{3}+\epsilon \frac{2(1+\gamma^{2})}{(1+u)\gamma}\frac{1}{R}\arctan{(\gamma R)}
\end{equation}
and
\bea
f(R)=\left(-\frac{\mu}{\sqrt{2|\beta|}}\frac{1}{R}+1 -\frac{R^{2}}{3}\right)\left( 1-\frac{2\epsilon}{(3 u-1)-(1+u)R^{2}} \right)-\frac{2\epsilon}{1+u} \nonumber \\ 
+ \frac{2( 1+\gamma^{2})}{(1+u)\gamma}\frac{1}{R}\arctan{(\gamma R)}
\eea
As long as our approximation $\epsilon<<|u-1|$ is true the linear perturbative solution is valid for large enough $R$ and the solution is regular and very similar to a de Sitter Schwarzschild solution. Note that the correction in $\epsilon$ does not interfere with the effective cosmological constant which means that the self-tuning mechanism is stable under a small deviation of the coupling $q$.
Alternatively for $u \in (-\infty ,-1)\cup (1/3, +\infty )$ we can set,
\begin{equation}
\omega=i \gamma \equiv \sqrt{\frac{1+u}{3 u-1}}
\end{equation}
to get,
\begin{equation}
h(r)=\left( 1-\frac{2}{1+u}\epsilon \right) -\frac{\mu}{R}-\frac{R^{2}}{3}-\epsilon \left(\frac{\omega^{2}-1}{\omega}\right)\frac{1}{(1+u)R} \ln{\left|\frac{ 1 +\omega R}{1 -\omega R}\right|}
\end{equation}
where we note now that the solution is again regular and the only difference is the slightly differing behavior of the logarithmic term.

\section{Concluding remarks}

In this paper we have considered a resolution method originally found in \cite{Babichev:2013cya} giving regular black holes for scalar tensor theories. Following the analysis in \cite{Babichev:2013cya} we showed how the complex system of field equations boiled down to solving a single third order algebraic equation (\ref{k}) which basically fixed the asymptotic properties of the spacetime solution. We then presented some new solutions analyzing in some detail the algebraic equation at hand (\ref{k}) which depends on two integration constants $C_0$ and $q$ (but not the mass of the black hole). The two integration constants are the fake hair charge $q$ and an integration constant $C_0$ determining the asymptotic properties of the solution at hand. Apart from novel regular black holes we also found that the particular de Sitter self tuning solution (\ref{desitter}) found in \cite{Babichev:2013cya} which is a self tuning black hole solution, is not some isolated point in the space of algebraic solutions, but is a generic and stable configuration with respect to small deviations in the fixed parameters of the solution. Indeed when we consider a small deviation in the self-tuning integration constant $q$ the effective cosmological constant of the neighboring black hole solution remains the same not spoiling the self tuning mechanism.   

The regularity and integrability of the method presented relies on three properties for the Galileon scalar field: first of all it has to have translational invariance and this symmetry yields a conserved charge and current with a special form for the scalar field equations. Secondly, the scalar tensor action considered has to include higher order Galileon terms so that the scalar field equation admits novel, geometric branches of solutions apart from the ones that lead to trivial or singular solutions. This enables us to satisfy the no hair arguments present for Galileons \cite{chanicolis} while finding non trivial scalar tensor black holes (see also  \cite{chazhou}). The regularity requirements required in \cite{chanicolis} are still not sufficient to obtain regular solutions. We need a third ingredient property : the key to obtaining a regular scalar field at the horizon is to additionally allow a linear time dependence for the scalar field. Under these three important properties it is then relatively easy to find novel scalar tensor black hole solutions. It would be interesting to see how far can one generalize these principles to go to more general Horndeski theories (see \cite{Kobayashi} for progress in this direction) or again find hairy black holes with a genuine scalar charge. In the latter case hairy black holes with genuine scalar charge have been found in a bi scalar tensor theory \cite{Charmousis:2014zaa} that are seen to elegantly generalize the BBMB black hole. Another interesting open question lies in the stability of such scalar-tensor black holes as those discussed here.  

\section*{Acknowledgements}
We are very happy to thank Eugeny Babichev, Carlos Herdeiro and Tassos Petkou for discussions and support. CC thanks the theoretical Physics group of Thessaloniki for their hospitality during the course and completion of this work. CC finally thanks colleagues from the University of Valencia for the invitation to give a talk on this subject and for the organisation of a very nice and interesting ERE 2014 conference.

\end{document}